\documentclass[letter, longauth]{aa}
\newcommand{\hii}{H\,{\scriptsize II}}

\usepackage[pdftex]{epsfig,graphicx,color}
\usepackage{txfonts}
\begin{document}
   \title{The \emph{Herschel}\thanks{Herschel is an ESA space
observatory with science instruments provided by European-led Principal
Investigator consortia and with important participation from NASA.} 
view of star formation in the Rosette molecular cloud under the influence of NGC~2244\thanks{
Figs~\ref{col}--\ref{center} are only available in electronic form at
http://www.aanda.org.}}

  \author{N. Schneider\inst{1},
          F. Motte\inst{1},
	  S. Bontemps\inst{1}\fnmsep\inst{2},
	  M. Hennemann\inst{1},
	  J. Di~Francesco\inst{3},  
          Ph. Andr\'e\inst{1}, 
          A. Zavagno\inst{4},     
          T. Csengeri\inst{1},  
          A. Men'shchikov\inst{1},\\
	  %
	  % and \\
          A. Abergel\inst{5},   
	  J.-P. Baluteau\inst{4},    
	  J.-Ph. Bernard\inst{6},   
	  P. Cox\inst{7},  
          %L. Deharveng\inst{4}, 
	  P. Didelon\inst{1},
	  A.-M. di Giorgio\inst{8},
          R. Gastaud\inst{1},  
	  M. Griffin\inst{9},       
	  P. Hargrave\inst{9},    
          T. Hill\inst{1},   
	  M. Huang\inst{10},        
	  J. Kirk\inst{9},         
          V. K\"onyves\inst{1}, 
	  S. Leeks\inst{11},        
	  J.Z. Li\inst{10},           
	  A. Marston\inst{12},       
	  P. Martin\inst{13},        
	  V. Minier\inst{1}, 
	  S. Molinari\inst{8},       
	  G. Olofsson\inst{14},     
	  %A. Omont\inst{15},  
          P. Panuzzo\inst{1},       
	  P. Persi\inst{15},        
	  S. Pezzuto\inst{8}, 
          H. Roussel\inst{16}, 
	  D. Russeil\inst{4},        
	  S. Sadavoy\inst{3},        
	  P. Saraceno\inst{8},      
	  M. Sauvage\inst{1}, 
	  B. Sibthorpe\inst{17},     
	  L. Spinoglio\inst{8},     
	  L. Testi\inst{18},        
	  D. Teyssier\inst{12},     
	  R. Vavrek\inst{12},       
	  D. Ward-Thompson\inst{9}, 
	  G. White\inst{11}\fnmsep\inst{21},         
	  C. D. Wilson\inst{19},       
	  A. Woodcraft\inst{20} \\  
          }
             %1
  \institute{Laboratoire AIM, CEA/DSM - INSU/CNRS - Universit\'e Paris 
        Diderot, IRFU/SAp CEA-Saclay, 91191 Gif-sur-Yvette, France 
%   \email{nschneid@cea.fr} 
            \and  %3 -> 2
             Laboratoire d'Astrophysique de Bordeaux, CNRS/INSU -- Universit\'e de Bordeaux, 
             BP 89, 33271 Floirac cedex, France
             \and  %5 -> 3 
             National Research Council of Canada, Herzberg Institute of Astrophysics,  
             University of Victoria, Department of Physics and Astronomy, Victoria, Canada 
            \and  %2 -> 5-> 4  
             Laboratoire d'Astrophysique de Marseille, CNRS/INSU - Universit\'e de Provence, 
             13388 Marseille cedex 13, France
             \and  %4 -> 5 
             IAS, Universit\'e Paris-Sud, 91435 Orsay, France 
             \and  %8 -> 4 -> 6 
             CESR \& UMR 5187 du CNRS/Universit\'e de Toulouse, BP 4346, 31028 Toulouse Cedex 4, France 
             \and %11 ->8 -> 7  
             IRAM, 300 rue de la Piscine, Domaine Universitaire, 38406 Saint Martin d'H\`eres, France   
             \and  %6 -> 8 
             INAF-IFSI, Fosso del Cavaliere 100, 00133 Roma, Italy
             \and  %7 -> 9
             Cardiff University School of Physics and Astronomy, UK
             \and %12 -> 11 -> 10 
              National Astronomical Observatories, Chinese Academy of Sciences, Beijing 100012, China
             \and %13 ->12 ->11
              Space Science and Technology Department, Rutherford Appleton Laboratory, Didcot, Oxon OX11 0QX, UK 
              \and %14 -> 13 -> 12
              Herschel Science Centre, ESAC, ESA, PO Box 78, Villanueva de la Ca\~nada, 28691 Madrid, Spain
              \and %15 _14 -> 13
              CITA \& Dep. of Astronomy and Astrophysics, University of Toronto, Toronto, Canada
              \and %16 -> 15 -> 14
              Department of Astronomy, Stockholm University, AlbaNova University Center, Roslagstullsbacken 21, 
              10691 Stockholm, Sweden
              \and %17 -> 16 -> 15
              INAF-IASF, Sez. di Roma, via Fosso del Cavaliere 100, 00133 Roma, Italy 
              \and 
              UPMC Universit\'e de Paris 06, UMR7095, IAP, 75014 Paris, France
              \and %18 -> 17 ->16
              Astronomy Technology Centre, Royal Observatory Edinburgh, Blackford Hill, EH9 3HJ, UK 
              \and %19 -> 18 -> 17
              ESO, Karl Schwarzschild Str. 2, 85748, Garching, Germany
              \and %20 -> 18
              Department of Physics and Astronomy, McMaster University, Hamilton, Canada 
              \and %21 -> 29
              SUPA, Institute for Astronomy, Edinburgh University, Blackford Hill, Edinburgh EH9 3HJ, UK
              \and 
              Department of Physics \& Astronomy, The Open University, Milton Keynes MK7 6AA, UK} 

\date{Received 2010 March 31st ; accepted }

\offprints{N. Schneider}

\mail{nschneid@cea.fr}

\titlerunning{The Herschel view of star formation in Rosette}

\authorrunning{N.~Schneider et al.}

\date{\today}

\abstract
% context 
{The Rosette molecular cloud is  promoted as the archetype of a triggered  
star-formation site. This is mainly due to its morphology, because the central 
OB cluster NGC~2244 has blown a circular-shaped cavity into the cloud and the 
expanding \hii-region now interacts with the cloud. } 
% aims 
{Studying the spatial distribution of the different evolutionary
states of all star-forming sites in Rosette and investigating possible
gradients of the dust temperature will help to test 
the 'triggered star-formation' scenario in Rosette.} 
% a long-standing question that has not yet been fully answered.
% methods
{We use continuum data obtained with the PACS (70 and 160 $\mu$m) and
SPIRE instruments (250, 350, 500 $\mu$m) of the
\emph{Herschel} telescope during the Science Demonstration Phase of HOBYS.}
% results
{Three-color images of Rosette impressively show how the molecular gas is heated 
by the radiative impact of the NGC~2244 cluster. A clear negative temperature gradient 
and a positive density gradient (running from the \hii-region/molecular cloud interface   
into the cloud) are detected. Studying the spatial distribution of the most massive dense cores 
(size scale 0.05 to 0.3 pc), we find an age-sequence (from more evolved to younger) with 
increasing distance to the cluster NGC~2244. No clear gradient is found for the clump 
(size-scale up to 1 pc) distribution.}  
% conclusions
{The existence of temperature and density gradients and the observed age-sequence  
imply that star formation in Rosette may indeed be influenced by the radiative impact 
of the central NGC~2244 cluster. A more complete overview of the prestellar and 
protostellar population in Rosette is required to obtain a firmer result. }
% 5 {} token are mandatory


   \maketitle
%
%________________________________________________________________

\begin{figure}[ht]
\includegraphics[angle=0,width=85mm]{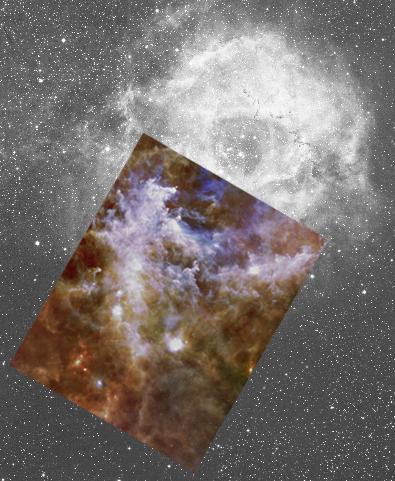}
\caption [] {Three-color image (70 $\mu$m = blue, 160 $\mu$m = green, 500  
$\mu$m = red) of Rosette, overlaid on an optical image (H$_\alpha$ from the Digital Sky Survey).}
\label{rosette-3col}
\end{figure}

\section{Introduction} \label{intro}
\begin{figure}[ht]
\includegraphics[angle=0,width=90mm]{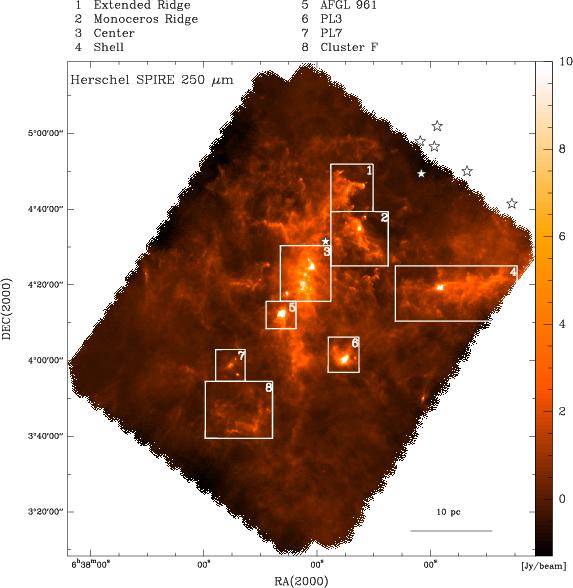}
\caption [] {\emph{Herschel}/SPIRE map at 250 $\mu$m of Rosette. Individual regions referred 
to in the text are numbered and labeled. The stars indicate the seven O-stars (O4 to O9) 
of NGC~2244.} 
\label{overview}
\end{figure}

The Rosette is a spectacular region due to its well-known emission
nebula, powered by the central OB cluster NGC~2244
(Figs.~\ref{rosette-3col}, \ref{overview}) that has blown a cavity
into the molecular cloud. The expanding \hii-region is interacting
with a high-mass (few times $10^5~M_\odot$) star-forming molecular
cloud at 1.6~kpc from the Sun (Williams et al. \cite{williams1994},
Schneider et al. \cite{schneider1998a}, Heyer et
al. \cite{heyer2006}), creating photon dominated regions (PDRs) along
the interface (Schneider et al. \cite{schneider1998b}).  Bright PAH
emission, first seen in the IRAS bands by Cox et al. (\cite{cox1990}),
indicates the presence of warm dust. Near- and mid-IR surveys (Phelps
\& Lada \cite{phelps1997}, Li \& Smith \cite{li2005}, Balog et
al. \cite{balog2007}, Roman-Zuniga et al. \cite{roman2008}, Poulton et
al. \cite{poulton2008}) investigated the IR-population of Rosette in
greater detail .

%and showed that all young stellar clusters
%are embedded in massive molecular clumps.

Ionizing radiation has an effect on the evolution of a molecular cloud
and is thus closely linked with star formation. Photoevaporation of
gas on an existing (primordial) clumpy cloud structure causes the
formation of {\sl globules}. Well-known features in the Rosette
molecular cloud (hereafter RMC) are the 'elephant trunks', 'speck'-
and 'teardrop' globules (Schneps et al. \cite{schneps1980}, White et
al. \cite{white1997}), which show bright rim emission. These globules can
form low-mass stars and shrink slowly (e.g. Lefloch \& Lazareff \cite{lefloch1994}).
%(e.g. Reipurth \cite{reipurth1983}, Lefloch \& Lazareff
%\cite{lefloch1994}, Sugitani et al. \cite{sugitani1994}).
Rayleigh-Taylor instabilities 
%(see, e.g., Woodward \cite{woodward1976}) 
can explain the elephant trunk globules in
Rosette (e.g. Patel et al. \cite{patel1993}).  
%
%A more global view of the
%effects of radiation on molecular clouds was given by, e.g.,  
%Bertoldi (\cite{bertoldi1989}) and 
%Bertoldi \& McKee (\cite{bertoldi1990}). In their view, 'radiation driven 
%implosion' can cause individual clumps to collapse and form stars within a 
%few free-fall time scales.
% or can lead to the formation of an 'equilibrium cometary
%cloud' without star-formation that photoevaporates slowly. 
Schneider et al. (\cite{schneider1998a}) 
showed that basically all CO clumps close to the \hii-region collapse
and form stars. Recently, Gritschneder et al. (\cite{grit2009}) ran hydrodynamic
models including turbulence and ionizing radiation. They showed that
it is due to the heating and consequently the increasing pressure of the
lower-density interclump gas that primordial, denser clumps are even
more compressed,  forming pillar-like structures that point towards the
source of radiation. In the high-density (n$\sim$10$^7$ cm$^{-3}$) tip
of the pillars, star formation takes place. However, the extent to which 
star formation is promoted deeper within the cloud, where the level of
ionization is drastically falling, is not clear.

%A typical feature of these is a velocity shift 
%of around 5 km/s compared to the bulk emission of the cloud. 
%
%cluster age 2 x 10^6 yr, v (ionization front = 10 or 20 km/s)
% -> for 10 km/s ionization front entered 10 pc, 
% -> for 20 km/s     ``                   23 pc 
%
% distance HII-region/PL7 is 23 pc

Although there is a clear influence of radiation on the evolution of the
RMC in general, it is less clear what its specific link to star
formation is. In the classical picture of {\sl sequential triggered} star
formation (Elmegreen \& Lada \cite{elmlada1977}), an expanding
\hii-region assembles gas in the layer compressed between the
ionization front and the shock front, which then fragments into
gravitationally unstable dense cores that form stars. The ionization
front of this new association propagates further into the cloud and
induces the next generation of star formation.

%Dale et al. \cite{dale2007} show that ionizing radiation can enhance 
%the formation of cores in a gloabbly unbound molecular cloud. 

%CHECK genrally: ionizing radiation disrupting a cloud or haveing a positive 
% feedback on SF ? 

The Rosette was proposed as an archetype of a triggered star
formation site. Many other studies (e.g. Deharveng et
al. \cite{deharveng2005}) have revealed possible candidates on smaller size
scales with a clearly-defined, simple geometry and indications of
star-formation in a swept-up shell of gas around a central
\hii-region.  The Rosette is a more demanding region, because it is a
massive molecular cloud with a complex structure caused by the 
presence of at least three clusters that illuminate the
\hii-region (NGC~2244, NGC~2237, and the recently detected RMC-XA 
(Wang et al. \cite{wang2009}). There are arguments
that star-formation in Rosette may be the result of a triggering
process: \\ (i) Balog et al. (\cite{balog2007}) and Roman-Zuniga et
al. (2008) found that the average NIR 
excess fraction for stellar clusters {\sl increases} with distance to
the cluster center. 
%
%Signs of 'sequential' star formation ? 
%that the ratio between
%Class I to Class II sources in NGC2244 is rather low (7\%), much lower
%than ratios observed in other star forming regions with smaller 
%proportions of massive stars. 
%(2) Heyer et al. (2006) detect a more turbulent velocity structure in
%the outer molecular cloud regions. A link to star formation, however,
%is not established.
%
(ii) CO line studies of Williams et al. (\cite{williams1994}) show
that star-formation activity is more intense in the
\hii-region/molecular cloud interface region than in the molecular
cloud center.  

%Dent et al. (\cite{dent2009}) detected a decreasing
%(into the cloud) surface temperature of the clumps due to a decreasing
%$^{12}$CO 3$\to$2/1$\to$0 line ratio.
%However, molecular line follow-ups of the clumps containing
%clusters (Roman-Zuniga et al. \cite{roman2008}) did not detect a
%variation of their physical properties (density, temperature etc.)
%across the cloud. 

%These arguments support the idea of {\sl triggered} star formation,
%but probably not in the original meaning of being {\sl sequential}.
However, the triggered star-formation scenario is likely not {\sl sequential}.
Roman-Zuniga et al. (\cite{roman2008}) point out that the relative age
differences of the clusters are not consistent with a sequential
triggered star formation scenario. They propose that the overall age
sequence of cluster formation could be primordial, possibly resulting
from the formation and evolution of the molecular cloud itself.  It is
thus unclear whether the star-formation activity in Rosette is
progressing from the \hii-region interface into the cloud center, or 
whether it is a global process with a random spread of evolutionary stages and no
obvious link between more and less evolved star-forming sites.

\section {Observations}  \label{obs}

The RMC was chosen as a representative source (in terms of mass,
star-formation activity etc.) for the HOBYS (\emph{Herschel} imaging
survey of OB Young Stellar objects) guaranteed time key program (Motte
et al. 2010) on the \emph{Herschel} satellite (Pilbratt et al. 2010),
using the SPIRE (Griffin et al. 2010, Swinyard et
al. (\cite{swinyard2010}), and PACS (Poglitch et al. 2010) continuum
arrays.  The SPIRE and PACS data from 70 to 500 $\mu$m were obtained
on october 20th 2009 in the parallel mode with a scanning speed of
20$''$/sec. Two cross-linked coverages of size
1$^\circ$45$'\times$1$^\circ$25$'$ were performed, focussing on the
southeast corner of the Rosette nebula, i.e. the main part of the
molecular cloud (see Fig.~\ref{overview}) The SPIRE data were reduced
with HIPE 2.0, using modified versions of the pipeline scripts (for
example observations that were taken during the turnaround at the map
borders were included). A median baseline and the `naive-mapper'
(i.e. a simple averaging algorithm) were applied to the data.

%The angular resolution at 70, 160, 250, 350, and 500 $\mu$m, is $\sim$6$''$,
%$\sim$12$''$, $\sim$18$''$, $\sim$25$''$, and $\sim$37$''$, respectively.}

%Two cross-linked coverages of size 1$^\circ$45$'\times$1$^\circ$25$'$
%were performed, focussing on the south-east corner of the Rosette
%nebula, i.e. the main part of the molecular cloud (see
%Fig.~\ref{overview}). The SPIRE data were reduced using a HIPE version
%2.0 and modified (e.g. observations that were taken during the
%turnaround at the map borders were included) pipeline scripts that
%were delivered with this version were used. A median baseline (HIPE
%default) was applied to the maps and as map-making algorithms, the
%'naive mapper' (HIPE default) was used.

%The resulting rms of the map is around 0.14 Jy/beam (250 $\micron$) 

%The final maps show a displacement of a $\sim$6 arcsec in RA and Dec 
%when compared to point sources from 2MASS. 

% 250 micron rms around 0.14 Jy/beam

\begin{figure}[ht]
\includegraphics[angle=-90,width=90mm]{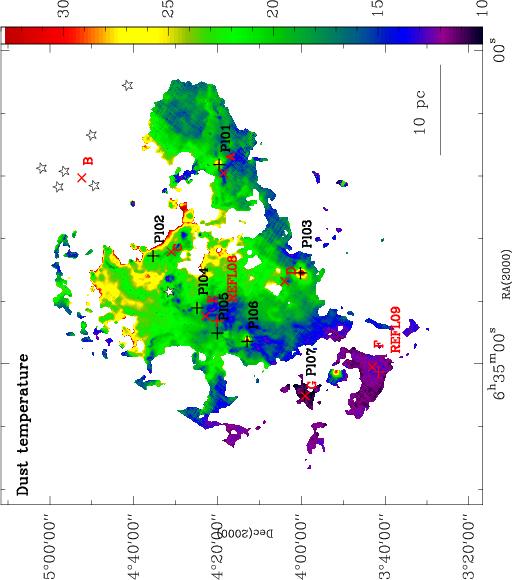}
\caption [] {Dust-temperature map of the Rosette molecular cloud,
obtained from simultaneously fitting the five observing wavebands of PACS
and SPIRE. The temperature scaling
in Kelvin is given on the right. Note that there is probably a
zero-point offset in the maps due to the median filter on the scan
legs.  Only areas that are covered by SPIRE {\sl and} PACS are
shown. Black crosses and label 'PL' indicate the position of the
clusters identified by Phelps \& Lada (\cite{phelps1997}), red crosses
and label 'A','B' etc. the clusters from Poulton et
al. (\cite{poulton2008}), and label 'REFL' the clusters from Roman-Zuniga
et al. (2008)}.
\label{temp}
\end{figure}

\section{Results and analysis} \label{results}

\subsection{Three-color image } 

Figure~\ref{rosette-3col} displays a three-color composite (70, 160,
and 500 $\mu$m) image of the RMC, overlaid on a H$\alpha$ image of
the region. Blue represents the shortest wavelength (70 $\mu$m) and
thus the warmer gas phase, green the 160 $\mu$m emission, and red the
coldest gas phase, traced by 500 $\mu$m emission.  The
\hii-region/molecular cloud interface, where the UV radiation has the
largest impact, is more prominent in blue and shows the most complex
and filamentary structure. There, the UV radiation from the NGC~2244
cluster (the 7 O- and 24 B-stars induce a streaming velocity at the
ionization front of up to 20 km s$^{-1}$ (Fountain et
al. \cite{fountain1979})) has erased the low-density gas, leaving only the
densest gas as `pillars'.  Assuming a cluster age of 2$\times$10$^6$
yr (Wang et al. \cite{wang2009}) and a typical expansion velocity of
the ionization front of 10 km s$^{-1}$, this would imply that the radiation 
penetrates to a depth of 
%a penetration depth of the radiation of 
$\sim$10 pc. This is indeed approximately
the border between the warm and colder gas phase.

Penetrating deeper into the cloud, the cold and dense `molecular
ridge' (mostly appearing in pink/red) is less influenced by UV
radiation and appears less structured. This does not imply that the
gas there has a smoother structure because (i) the lower-density
gas still present in this part of the cloud smoothes the distribution, and
(ii) resolution effects may play a role.
% (the dominating `red' 500
%$\mu$m emission has a resolution of $\sim$37$''$ in contrast to the 70
%$\mu$m emission with 6$''$). 
The `coldest' region is found in a 
remote part of the cloud at 20--30 pc, close to cluster PL7 (see
Fig.~\ref{overview}) and beyond.

%Summarizing, the spatial structure detected by the dust emission at
%the different wavelengths does not show a clear spatial dependence
%(for example a more complex structure close to the \hii-region) that
%would point towards an influence of the OB cluster. Moreover, the
%cloud reflects the typical clumpy turbulent structure characteristic
%for other low- and high-mass star forming molecular clouds and may have
%been generated by other physical processes (cloud formation due to
%colliding flows, energy injection by supernova-explosions etc.).

%	beta  = 2.				; dust emissivity index
%	kappa1000 = 0.1				; cm^2/g  (at 1000 GHz)
%	kappa  =  kappa1000*(freq/1.e3)^beta	; cm^2/g  (at freq)

\subsection{Dust temperature and column density} 

% dust temperature map 
Figure~\ref{temp} shows the distribution of the dust
temperature obtained from a greybody fit to the five observed
wavebands at each pixel (all maps were smoothed to the beamsize of
the 500 $\mu$m map, i.e. $\sim$37$''$). A dust emissivity index of
$\beta$=2 and a dust opacity 
%(mass absorption coefficent per gramm dust) 
of $\kappa=0.1*(\nu/1000)^\beta$ [cm$^2$/g] were used. 
%
%Due to the uncertainties in the flux calibration, the absolute values
%have an accuracy of about 15--20\%.
%
The highest temperatures are found at the
\hii-region/molecular cloud interface at the Monoceros Ridge and Shell
region, indicated by a thin layer of temperatures around 30
K. The temperature then gradually decreases into the molecular cloud 
down to around 10--15~K in the remote part of the cloud. This is
consistent with the dust color temperatures derived with 
the IRAS 60/100 $\mu$m ratio (Schneider et al. \cite{schneider1998b}),
which gives 30~K at the \hii-region/molecular cloud interface and 26~K
in the remote part of the cloud at the position of AFGL961. (Note that
the latter values are derived from low-angular resolution IRAS data.)
 
% in $\sim$30 pc distance.

%The UV field varies from 200 G$_\circ$ at the interface (15 pc distance from 
%the central OB cluster) to a few 10 G$_\circ$ in the remote cloud (30 pc).

% Dust properties 
The temperature gradient indicated by the PACS/SPIRE observations
confirms earlier findings by Cox et al. (1990), who showed that 60 and
100 $\mu$m IRAS emission is strong in the molecular cloud and nebula
region, while the 12 $\mu$m emission is prominent in the molecular
cloud {\sl and} inside the \hii-region. Dent et al. (\cite{dent2009})
detected a decreasing (into the cloud) surface temperature of the
clumps due to a decreasing $^{12}$CO 3$\to$2/1$\to$0 line ratio.
%
%This points towards a destruction of grains due to UV radiation, where
%only the 25 $\mu$m emission remains partly strong in the ionized
%regime, indicating a type of dust that is more resistant to
%radiation. The warm dust (12 $\mu$m IRAS and 70 $\mu$m emission of
%PACS) delineates a shell enclosing the ionization front. It is there
%where the \hii-region impacts the molecular cloud.
%
%Correlation with clusters 
Not all sites of star formation are correlated with temperature peaks.
Clusters A, C, D, E, F, G(=PL7) and PL5 do not appear prominent in the
temperature map. In contrast, one strong temperature peak in the
remote part of the cloud (between PL7 and F) is not identified as a
cluster.

From the column density map (Fig.~\ref{col}, right), derived from the
greybody fit, we calculated the total mass of the complex mapped with
\emph{Herschel} to be $\sim$10$^5$ M$_\odot$. The mass and average
density of the individual regions is given in
Table~\ref{cores}. Interestingly, there is a clear increasing gradient
in average {\sl density} from the \hii-region into the molecular
cloud. In the `compression' zone of the molecular cloud/\hii-region
interface (Shell/Extended and Monoceros Ridge), the densities are
around 0.5$\times$10$^3$ cm$^{-3}$, while the most remote clouds
(Cluster F and PL7) have a higher density of 1.8$\times$10$^3$
cm$^{-3}$ and 3.9$\times$10$^3$ cm$^{-3}$, respectively.

%Note that there is an offset of a few A$_v$ between the maps so that 
%for mass determinations, we corrected for the missing intensity. 

%However, the largest concentration of regions
%with high column-densities is found in the remote part of the cloud
%with the clumps containing cluster F and PL7. 

\subsection{A closer look at individual regions} \label{detail} 

\begin{figure}[ht]
\includegraphics[angle=0,width=90mm]{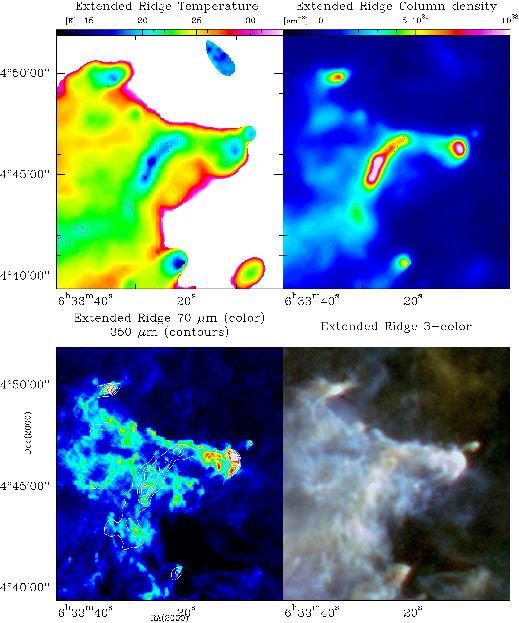}
\caption [] {Zoom into the {\sl Extended Ridge} region. Upper left: dust temperature, 
upper right: column density, lower left: Overlay 70 $\mu$m and 350 $\mu$m, lower 
right: three-color image  (70 $\mu$m = blue, 160 $\mu$m = green, 250 $\mu$m = red). 
Note that here we used the 250 $\mu$m data for the red band.} 
\label{extended}
\end{figure}

Following the notation given in Fig.~\ref{overview}, we show the {\sl
Extended Ridge} in Rosette in more detail. Other regions (the {\sl
Center} and the {\sl Monoceros Ridge}) can be found as online
figures. The {\sl Extended Ridge} represents dense, molecular pillars
that are exposed to the ionizing radiation from the NGC~2244
cluster. There are strong temperature and density gradients, visible
in the maps of dust temperature, column density, and in the
three-color plot shown in Fig.~\ref{extended}. The region of highest
column density forms a chain of cold, dense gas clumps, as can be seen
in the overlay between 70 $\mu$m and 350 $\mu$m emission in
Fig.~\ref{extended} where the coldest regions do not appear in 70
$\mu$m emission. The masses of these clumps range between 19 and 161
M$_\odot$, and their average density is typically 1.1$\times$10$^4$
cm$^{-3}$ (values determined from the column density map).
%Compare with Dent et al. 2009, they obtain 3-100 Msun for the clumps 
%in the expanding ring.
Though this average density is not low, it is far from the densities
of 10$^7$ cm$^{-3}$ predicted by turbulence models for pillar
formation (Gritschneder et al. \cite{grit2009}).

%However, Patel et al. (\cite{patel1993}) and Schneider et al. (1998a)
%showed that the globules in the SE part of the molcular cloud cannot
%be formed as Rayleigh-Taylor instabilities but by radiation driven
%implosion.

%D-type ionization fronts inside massive clumps cause the
%formation of globules.

%D-type: subsonic acceleration of neutral gas 
%R-type: supersonic acceleration 

\section{Discussion} \label{discussion} 

The prestellar and protostellar population of Rosette observed with
\emph{Herschel} was identified with two source extraction techniques
that are both based on the principle that emission on large spatial
scales is filtered out in order to focus on compact sources.
Motte et al. (2010) discuss the core population 
(prestellar, protostellar, and warm cores on a size scale 0.05--0.3 pc).
%For the compact (0.05--0.2 pc) sources, we will use the
%results from the {\sl Gaussclumps} technique, and for the clumps (up
%to 1 pc), the {\sl Getsources} results that are described in 
Di~Francesco et al. (2010) present the {\sl clump} distribution (up to
1 pc), which is either starless or associated with YSO.

%Both papers explain how they define the evolutionary status of different sources.  

%\subsection{Spatial distribution of clumps and cores in different evolutionary  stages} 

%\noindent {\bf Clumps} \\ 
Figure~1 in Di~Francesco et al. (2010) shows the spatial distribution
of starless and protostellar {\bf clumps}. The latter are concentrated in
the densest cloud regions, mainly close to the known star clusters,
while the starless clumps are more evenly distributed.  {\sl No}
correlation between clump type and distance to the OB cluster is
observed that may suggest an age-effect. 
 
%\noindent {\bf Dense cores} \\ 
%In Fig.~4 in Motte et al. (2010), the spatial distribution of the most
%massive dense {\bf cores} is shown and in Table~1 (this paper), the
%statistics of these sources with regard to the different regions is
%summarized.  If we assume that the rough classification that Motte et
%al. (2010) have done for these 0.1 pc dense cores is indeed showing an
%evolutionary sequence from prestellar cores to evolved YSO, we can
%compute the relative numbers of young vs. evolved sources in each
%region (last column in Table~1).  The resulting sequence suggests that
%there is indeed an age-gradient across the RMC with the PL7 cluster
%being the youngest one in the remote part of the cloud and the
%Monoceros Ridge and Shell Region close to the \hii-region being the
%most evolved. 

In Fig.~5 in Motte et al. (2010) the spatial distribution of the
most massive dense {\bf cores} is shown and in our Table~1 the
statistics of these sources with regard to the different regions is
summarized.  We adopt the classification for prestellar cores to
evolved YSO defined in Motte et al. (2010) and compute the relative
numbers of young vs. evolved sources in each region (last column in
Table~1). The resulting sequence {\sl may} indicate that there is an
age-gradient across the RMC with the PL7 cluster being the youngest,
in the remote part of the cloud, and the Monoceros Ridge and Shell
Region, close to the \hii-region, being the most evolved. An
independent confirmation of the youth of cluster PL7 is given in
Hennemann et al. (2010) in their protostellar envelope mass
vs. bolometric luminosity diagram. Note that Roman-Zuniga et
al. (2008) also detected an increasing fraction of young stars with
increasing distance to the NGC~2244 cluster center and Balog et
al. (\cite{balog2007}) found that the Class II sources are
concentrated at the center of NGC~2244, while the Class I sources are
located further away from the cluster. On the other hand, cluster PL3 does
not fit this scenario well, because it seems to be a more evolved region,
but is located in the more remote part of the cloud. A tentative
explanation may be that because it is more exposed to UV radiation and
thus less shielded by denser gas, (like the Center and remote region)
its evolution might have been accelerated by the radiative impact.
However, we emphasize that there are not enough sources analyzed so
far to draw a firm conclusion. A more detailed analysis with a larger
sample and significant statistics is required, which is planned for the
future. 

%The disk ratio increases from 27\% within the
%0.5 pc up to 45\% within 0.5 to 2.5 pc. They found a strong decrease
%in the photoevaporation rate with distance to the cluster.

%\begin{table}[htbp]
% \centering
%   \begin{tabular}{lcc}
%\hline
%\hline
%                             & $<$n$>$               & M            \\
%Region                       & [10$^3$ cm$^{-03}$]   & [M$_\odot$]  \\ 
%\hline                        
%4   Shell                    &  0.4                  & 11780 \\ 
%1+2 Extended+Mon. Ridge      &  0.6                  &  4280 \\
%3 Center                     &  0.6                  & 22250 \\
%6 PL3                        &  1.5                  &  1650 \\
%8 Cluster F/REFL09           &  1.8                  &  5420 \\
%7 PL7                        &  3.9                  &  2080 \\
%\end{tabular}
%\caption{Average density  ($<$n$>$) and mass (M) derived from the column density map in the different regions 
%of the RMC (see Fig.~4), ordered by density.} 
%\label{masses}
%\end{table}

\begin{table}[htbp]
% \centering
   \begin{tabular}{lcccc}
\hline
\hline
                        & d    & $<$n$>$ & $M$      &  Young/ \\
Region                  & [pc] &  [10$^3$ cm$^{-03}$]  &  [M$_\odot$] &  evolved \\ 
\hline
                        &      &         &        &    \\                       
7 PL7                   &  35  &  3.9    &  2080  &  3/0 \\ % 3/0
6 PL3                   &  27  &  1.5    &  1650  &  1/3 \\ % 1/2 
5 AFGL961               &  25  &  9.5    &   792  &  4/1 \\ % 3/1
3 Center                &  20  &  0.6    & 22250  &  9/4 \\ % 8/5
1+2  Ext. + Mon. Ridge  &  10  &  0.6    &  4280  &  6/3 \\ % 6/3
4 Shell                 &  10  &  0.4    & 11780  &  1/2 \\ % 1/2
\end{tabular}
\caption{Age-sequence of the sources found in Rosette. The regions are
ordered with decreasing distance (d) from the center of NGC~2244 (for
simplicity we take as 'center' the position of cluster B in
Fig.~3). The average density ($<$n$>$) and mass ($M$) were derived from the
column density by determining the area of each region and assuming the same 
extent of the cloud along the line of sight. 
The last column gives the number of sources 
(young vs evolved).}
\label{cores}
\end{table}

\begin{acknowledgements}
SPIRE has been developed by a consortium of institutes led by Cardiff
University (UK) and including Univ. Lethbridge (Canada); NAOC (China);
CEA, LAM (France); IFSI, Univ. Padua (Italy); IAC (Spain); Stockholm
Observatory (Sweden); Imperial College London, RAL, UCL-MSSL, UKATC,
Univ. Sussex (UK); and Caltech, JPL, NHSC, Univ. Colorado (USA). This
development has been supported by national funding agencies: CSA
(Canada); NAOC (China); CEA, CNES, CNRS (France); ASI (Italy); MCINN
(Spain); SNSB (Sweden); STFC (UK); and NASA (USA).\\
%The following institutes have provided hardware and software elements
%to the SPIRE project: University of Lethbridge, Canada; NAOC, Beijing,
%China; CEA Saclay, CEA Grenoble and OAMP in France; IFSI, Rome, and
%University of Padua, Italy; IAC, Tenerife, Spain; Stockholm
%Observatory, Sweden; Cardiff University, Imperial College London,
%UCL-MSSL, STFC-RAL, UK ATC Edinburgh, and the University of Sussex in
%the UK.  Funding for SPIRE has been provided by the national agencies
%of the participating countries and by internal institute funding: CSA
%in Canada; NAOC in China; CNES, CNRS, and CEA in France; ASI in Italy;
%MEC in Spain; Stockholm Observatory in Sweden; STFC in the UK; and
%NASA in the USA.  Additional funding support for some instrument
%activities has been provided by ESA. \\
PACS has been developed by a consortium of institutes led by MPE
(Germany) and including UVIE (Austria); KU Leuven, CSL, IMEC (Belgium); CEA,
LAM (France); MPIA (Germany); INAF-IFSI/OAA/OAP/OAT, LENS, SISSA
(Italy); IAC (Spain). This development has been supported by the funding
agencies BMVIT (Austria), ESA-PRODEX (Belgium), CEA/CNES (France),
DLR (Germany), ASI/INAF (Italy), and CICYT/MCYT (Spain). \\
Part of this work was supported by the ANR (\emph{Agence Nationale  pour 
la Recherche}) project ``PROBeS'', number ANR-08-BLAN-0241.\\
The figures were prepared using GILDAS\footnote{http:iram.fr/IRAMFR/GILDAS} 
and with the help of the ESA/ESO/NASA Photoshop FITS Liberator. \\ 
We thank the referee, Marc Pound, for his critical comments. 
\end{acknowledgements}

\onlfig{5}
{
\begin{figure*}[ht]
\includegraphics[angle=-90,width=90mm]{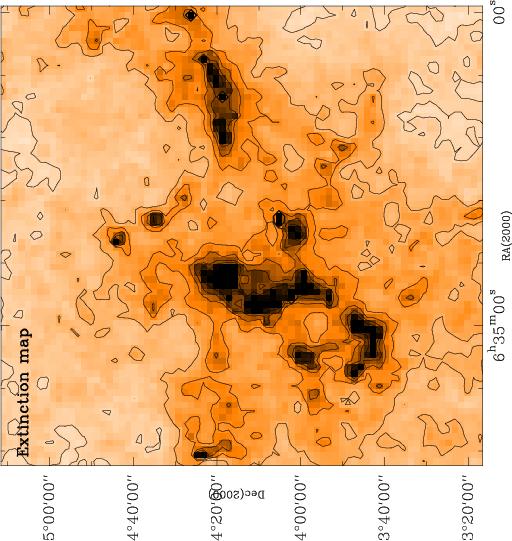}
\hskip 0cm
\includegraphics[angle=-90,width=88mm]{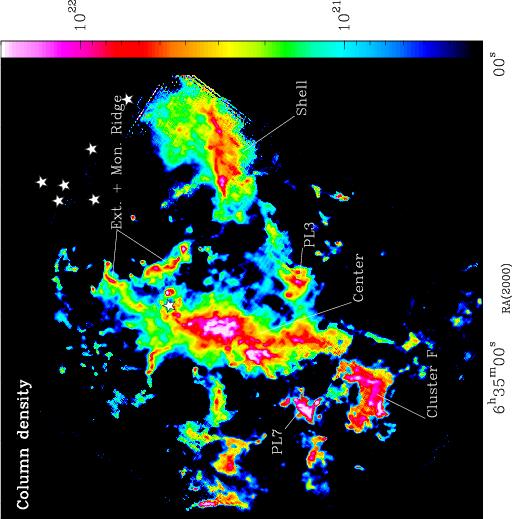}
\caption [] {{\bf Left:} Extinction map of the RMC, derived from
stellar reddening of background stars in JHK, using the 2MASS
database, and a pixel size of $1.3\arcmin$ (Bontemps, priv. comm.,
Schneider, Bontemps et al. \cite{schneider2010}). Contour levels increase from A$_V$=2$^m$ to
14$^m$ in steps of 2$^m$. {\bf Right:} Molecular hydrogen column
density [cm$^{-2}$] in logarithmic scaling determined from the same
greybody fit that was used for the temperature. The regions referred to in Table~1 are 
indicated.}
\label{col}
\end{figure*}
} 

\onlfig{6}
{
\begin{figure*}[]
\includegraphics[angle=0,width=90mm]{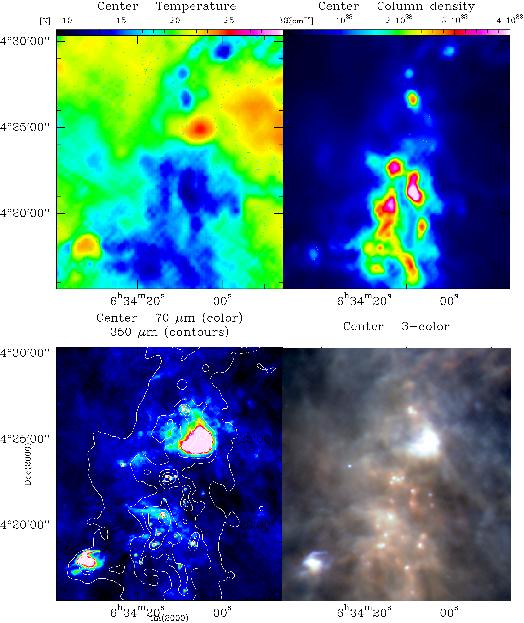}
\includegraphics[angle=0,width=90mm]{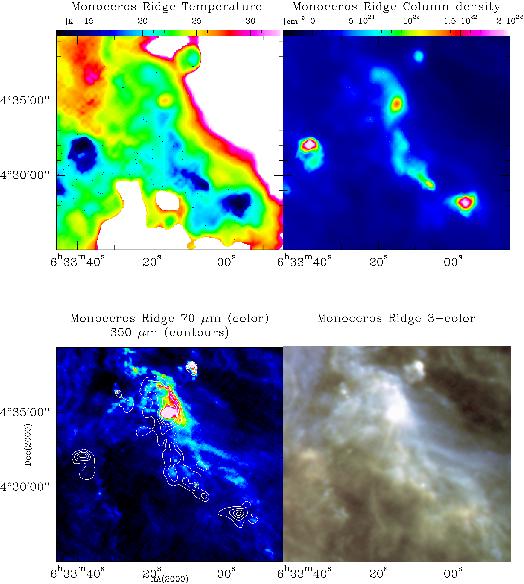}
\caption [] {Zoom into the {\sl Center} (left) and {\sl Monoceros
Ridge}(right) regions, similar to the Extended Ridge presented in
Fig. 4.Upper row: Dust temperature, and column density plots of each
region, respectively. lower row: Overlay 70 $\mu$m and 350 $\mu$m,
and three-color image (70, 160, 250 $\mu$m) of each region, 
respectively.}
%Zoom into the {\sl Center} and {\sl Monoceros Ridge} region. Upper left column: Dust temperature, 
%upper right column: column density, lower left: Overlay 70 $\mu$m and 350 $\mu$m, lower 
%right: three-color image (70, 160, 250 $\mu$m).} 
\label{center}
\end{figure*} 
}

%\onlfig{7}
%{
%\begin{figure}[]
%\includegraphics[angle=0,width=85mm]{mon_3_paper_final.jpg}
%\caption [] {Zoom into the {\sl Monoceros Ridge} region. Upper left: Dust temperature, 
%upper right: column density, lower left: Overlay 70 $\mu$m and 350 $\mu$m, lower 
%right: three-color image (70, 160, 250 $\mu$m) .} 
%\label{mon}
%\end{figure}
%} 

%\begin{eqnarray}
%N [cm^{-2}] & = & f(T_{ex}) \int T_{mb} [K] d{\rm v} [km s^{-1}]
%\nonumber \\ 
%\end{eqnarray}
%with 
%\begin{eqnarray}
%f(T_{ex}) & = & \frac{3hZ}{8 \pi^3 \mu^2 J_t}
%\frac{\exp(h\nu/kT_{ex})}{[1-\exp(-h\nu/kT_{ex})] (J(T_{ex}) - J(T_{BG}))} 
%\nonumber \\  
%& &   
%\label{ncol}  
%\end{eqnarray}

\end{document}